\documentclass[aps,prl,twocolumn,tightenlines,superscriptaddress]{revtex4}
\pdfoutput=1
\usepackage{epsfig}
\usepackage{graphicx}
\usepackage{psfrag}
\usepackage{amsmath,amssymb,psfrag,slashed,graphicx}
\usepackage{colordvi}
\usepackage{color}

\def \t {\tilde}

\begin{document}

\title{Quasi parton distribution functions at NNLO:  flavor non-diagonal  quark contributions }

\author{Long-Bin Chen$^1$, Wei Wang$^{2}$\footnote{Corresponding author:wei.wang@sjtu.edu.cn}, Ruilin Zhu$^{3,4}$\footnote{Corresponding author:rlzhu@njnu.edu.cn}}
\affiliation{
$^1$ School of Physics and Materials Science, Guangzhou University, Guangzhou 510006, China\\
$^2$ INPAC,  SKLPPC, MOE KLPPC, School of Physics and Astronomy, Shanghai Jiao Tong University, Shanghai, 200240,   China\\
$^3$ Department of Physics and Institute of Theoretical Physics,
Nanjing Normal University, Nanjing, Jiangsu 210023, China\\
$^4$ Nuclear Science Division, Lawrence Berkeley National
Laboratory, Berkeley, CA 94720, USA}

\date{\today}
\vspace{0.5in}
\begin{abstract}
We present a next-to-next-to-leading order (NNLO) calculation of the quasi parton distribution functions (Quasi-PDFs) in the large momentum effective theory (LaMET). We focus on the flavor non-diagonal quark-quark channel and demonstrate the LaMET factorization at the NNLO accuracy in the modified minimal subtraction scheme. The matching coefficient between the quasi-PDF and the light-cone PDF is derived. This provides a first step towards a complete NNLO analysis of quasi-PDFs and to better understand the nucleon structures from the  first principle of QCD.
\end{abstract}

\maketitle

Understanding the underlying  structure of nucleons from degrees of quarks and gluons  has been a long-standing  goal in hadron physics. Since deep-inelastic scattering experiments at Stanford Linear Accelerator Center in late 1960's, the proton structure has been explored in various hard
scattering processes~\cite{Gao:2013xoa}. The key results involve the parton distribution functions (PDFs), defined as momentum distributions of quarks and gluons in an infinite-momentum hadron. These distribution functions are normally referred as the light-cone PDFs or the collinear PDFs. In high energy experiments at the lepton-hadron and hadron-hadron colliders, the PDFs are also the important ingredients to characterize the structure of hadrons and make predictions for various processes to test the standard model and probe the new physics beyond. Though the scale evolution of PDFs beyond leading order (LO) into next-to-next-to-next-to leading order (NNNLO) have been performed in literatures~\cite{Furmanski:1980cm,Curci:1980uw,Moch:2002sn,Moch:2004pa,Vogt:2004mw},  calculating the PDFs and more generally  light-cone observables from first principle of  quantum chromodynamics (QCD),
has been extremely difficult.  In the formulation of non-perturbative QCD on a Euclidean lattice, one cannot directly explore  time-dependent correlations. Instead,  only  moments of parton distribution functions,  matrix elements of local operators, can be calculated.  However, the difficulty in   Lattice QCD study
grows significantly  for higher moments due to  technical reasons and thus  only limited moments can be  extracted to date~\cite{Martinelli:1987zd,Martinelli:1988xs,Detmold:2001dv,Dolgov:2002zm}.

An effective theory, called large momentum effective theory (LaMET)~\cite{Ji:2013dva,Ji:2014gla}, has been developed to compute various parton distribution functions on Lattice. In this framework, an appropriate static-operator matrix element (quasi-observable) that approaches the parton observable in the infinite momentum limit of the external hadron is constructed. The quasi-observable constructed in this way is usually hadron-momentum-dependent but time-independent, and thus can be readily computed on the lattice. After the renormalization, the  quasi-observable can be used to  extract the parton observable through a factorization formula accurate up to power corrections that are suppressed by the hadron momentum. The relevant parton distribution functions calculated in the LaMET are referred as Quasi-PDFs. Great progress has been made in the last few years on both the theoretical understanding of the formalism and the lattice simulations for parton distributions of baryons and mesons, see, for example, some recent reviews in Ref.~\cite{Cichy:2018mum,Ji:2020ect}.

The factorization arguments of LaMET allow us to carry out order by order perturbative calculations on the matching between the Quasi-PDFs and the light-cone PDFs. This matching is one of the crucial elements in applying LaMET to parton physics. It provides a solid foundation to compute the light-cone PDFs in a systematically controlled way. In some sense, the improvement on the precision of the PDF calculations can only be achieved by combining the advanced lattice simulations for the Quasi-PDFs (toward small lattice spacing, large volume and physical pion mass) and higher order perturbative matching calculations.

Higher order perturbative calculations are also important to demonstrate the factorization in the LaMET explicitly. In particular, some specific features of the factorization can only be manifest in the non-trivial two-loop calculations. Quasi-PDFs at one-loop order and the associated matching coefficients has been a subjective of active research since LaMET was proposed in 2013. This includes quark distribution~\cite{Xiong:2013bka,Ishikawa:2017faj,Izubuchi:2018srq,Ji:2018hvs}, gluon distribution~\cite{Wang:2017qyg,Wang:2017eel,Wang:2019tgg} and many others (See the review~\cite{Ji:2020ect}). The goal of this paper is to go beyond the one-loop order and perform, for the first time, a two-loop computation of the Quasi-PDF in the LaMET, taking the non-diagonal quark-quark channel as an example. This channel starts at two-loop order, which allows us to demonstrate the factorization in an intuitive method. We also notice that recently, the renormalization of Quasi-PDF operators have been studied at two loop order~\cite{Ji:2015jwa,Braun:2020ymy}. Together with this result, our paper will provide an important step toward a complete two-loop calculation of Quasi-PDF and the associated matching coefficients.

The rest of this letter is organized as follows. We first present our main result of non-diagonal quark-quark splitting in LaMET at two-loop order. We will provide a detailed calculations and demonstrate the factorization in detail. Based on these results, we show the matching coefficients at this order. Since the non-diagonal quark-quark splitting only starts at two-loop order, this presents the leading contribution for this channel. Some numeric results will also be presented to illustrate the behavior of the matching coefficients. We will summarize our work in the end.

 \textit{Factorization at Two-loop Order.}---
We start with the definitions of the light-cone PDF and Quasi-PDF. For the  unpolarized quark light-cone PDF, we have
\begin{align}\label{eq:pdf}
f_{q/H}(x,\mu)&=\!\int\!\! \frac{d\xi^-}{4\pi} \, e^{-ixp^+\xi^-}
\! \big\langle p \big| \bar{q}(\xi^-) \gamma^+ \nonumber\\&~\times
\exp\bigg(-ig \int_0^{\xi^-}d\eta^- A^+(\eta^-) \bigg)
q(0) \big|p\big\rangle,
\end{align}
where  $x= k^+/p^+$ is the quark longitudinal momentum fraction and $p^\mu=(p^0,0,0,p^z)$ is the hadron momentum. Similarly, the quasi-PDF for the unpolarized quark is defined as
\begin{align}\label{eq:quasipdf}
{ \tilde  f}_{q/H}(y, p^z) &= N\int \frac{dz}{4\pi} e^{iz  yp^z}  \langle p|\overline{q} (z)
   \Gamma \nonumber\\&~~\times\exp\bigg(-ig \int_0^{z}dz' A^z(z') \bigg) q(0) |p\rangle,~
\end{align}
where $z$ is a spatial direction and we will adopt $\Gamma=\gamma^t$ with the normalization factor $N= p^z/p^t$ and use the $\slashed{p}$ projector.

According to the factorization in the LaMET, we can write down the Quasi-PDFs $\tilde f_{q/H}(y ,p^z)$ in terms of the light-cone PDFs $f_{q'/H}(x, \mu)$:
\begin{align}\label{allfac}
\tilde f_{q/H}(y ,p^z)=&\int_{-1}^1\frac{dx}{|x|}\Big[C_{qq'}\Big(\frac{y}{x}, \frac{|x| p^z}{\mu} \Big)f_{q'/H}(x, \mu) \Big],
\end{align}
where  $q', q$ being the partons in the hadron. The $\tilde f_{q/H}(y ,p^z)$ is  an equal-time correlation while $f_{q'/H}(x, \mu) $ is lightcone PDF.   Though  $\tilde f_{q/H}(y ,p^z)$  and $f_{q'/H}(x, \mu) $ share the same infrared structure, their ultraviolet behaviors are different, and embedded in the short-distance coefficient  $C_{qq'}$.

\begin{figure}[th]
\begin{center}
\includegraphics[width=0.4\textwidth]{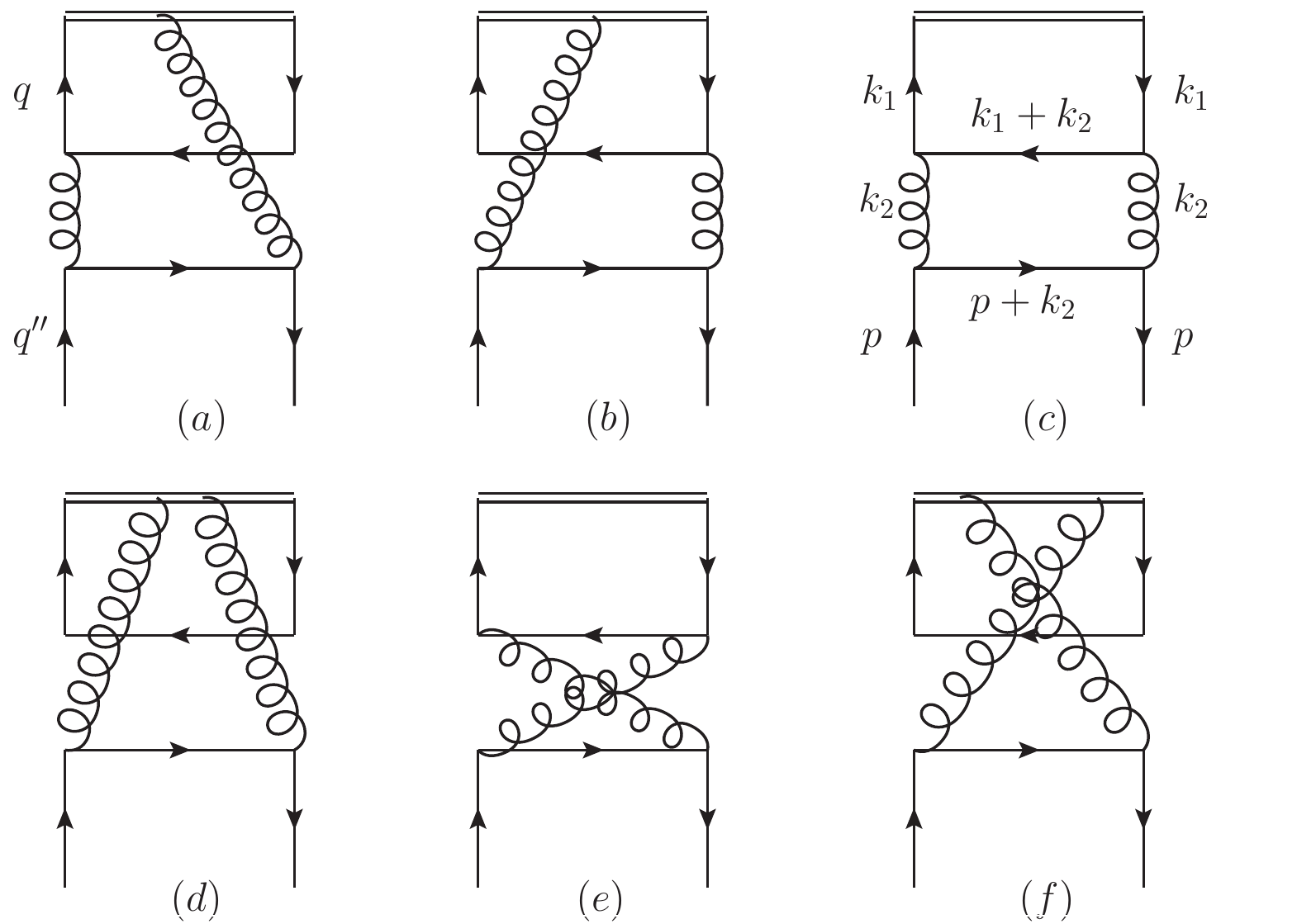}
\caption{ Feynman diagrams for $\t f_{q/q''}$ at NNLO, where $q$ and $q''$ are  quarks with  different flavours.  The double-lines correspond to  Wilson line.  }\label{tab:FDsea}
\end{center}
\end{figure}

Since the short-distance coefficient  is insensitive to the incoming hadrons,  in the calculation of  $C_{qq'}$  one can replace the hadron by the partonic state. In this work we will consider the flavor non-diagonal  quark contributions and the hadron state $|H\rangle$ is replaced by a quark state $|q''\rangle$ and we have the condition $q''\neq q$. We plot the Feynman diagrams for flavor non-diagonal  quark  distributions in Fig.~\ref{tab:FDsea}. In our computations below, we will apply the modified minimum subtraction scheme ($\overline{\rm MS}$) and dimensional regulation with $D=4-2\epsilon$. Under this scheme, we can write the formula for the flavor non-diagonal quark distribution as
\begin{align}\label{allfac}
\tilde f_{q/q''}(y ,\frac{ p^z}{\mu},\epsilon_{\mathrm{IR}})=&\int_{-1}^1\frac{dx}{|x|}\Big[C_{q q'}\Big(\frac{y}{x}, \frac{|x| p^z}{\mu} \Big)f_{q'/q''}(x, \epsilon_{\mathrm{IR}}) \Big]\nonumber\\=&~C_{q q'}\Big(\frac{y}{x}, \frac{|x| p^z}{\mu} \Big)
\otimes f_{q'/q''}(x, \epsilon_{\mathrm{IR}}) ,
\end{align}
where both sides are computed with dimensional regulations and $(1/\epsilon_{\rm IR})^n$ represent the Infrared divergences. At NNLO, the matching scheme is given as
\begin{align}
\t f^{(2)}_{q/q''}(y, \frac{ p^z}{\mu},\epsilon_{\mathrm{IR}})&=C^{(2)}_{q q'}\Big(\frac{y}{x}, \frac{|x| p^z}{\mu} \Big)\otimes f^{(0)}_{q'/q''}(x, \epsilon_{\mathrm{IR}})\nonumber\\&~+C^{(1)}_{q q'}\Big(\frac{y}{x}, \frac{|x| p^z}{\mu} \Big)\otimes f^{(1)}_{q'/q''}(x, \epsilon_{\mathrm{IR}})\nonumber\\&~+C^{(0)}_{q q'}\Big(\frac{y}{x}, \frac{|x| p^z}{\mu} \Big)\otimes f^{(2)}_{q'/q''}(x, \epsilon_{\mathrm{IR}}). \label{match.scheme}
\end{align}
Here, we have applied the perturbative expansions $T_i=\sum_{n=0}^\infty \left(\frac{\alpha_s}{2\pi}\right)^n T_i^{(n)}$ with $T_i$ being each of $\t f_{q/q''}, C_{qq'}, f_{q'/q''}$. Because of the particular feature of non-diagonal quark-quark splitting, each term at the right hand side of the above equation represents only one contribution. In the first term, $q'$ has to be $q^{\prime\prime}$, so that it only has $C_{q/q^{\prime\prime}}^{(2)}$. For the second term, $q'$ has to be a gluon, and the combination is quark-to-gluon splitting $f_{g/q}^{(1)}$ and gluon-to-quark $C_{qg}^{(1)}$ matching. Finally, $q'$ in the third term has to be $q$, representing non-diagonal quark-quark collinear splitting $f_{q/q''}^{(2)}$. We also know that both $f_{q''/q''}^{(0)}$ and $C_{q/q}^{(0)}$ are Delta functions. Therefore, the above equation can be simplified as
\begin{align}
\t f^{(2)}_{q/q''}(y, \frac{ p^z}{\mu},\epsilon_{\mathrm{IR}})=&C^{(1)}_{q g}\Big(\frac{y}{x}, \frac{|x| p^z}{\mu} \Big)\otimes f^{(1)}_{g/q''}(x, \epsilon_{\mathrm{IR}})\nonumber\\&+C^{(2)}_{q q''}\Big(y, \frac{ p^z}{\mu} \Big)+ f^{(2)}_{q/q''}(y, \epsilon_{\mathrm{IR}}) \ .\label{match.scheme.simple}
\end{align}
Here, $C^{(1)}_{q g}$, $f^{(1)}_{g/q''}$ and $f^{(2)}_{q/q''}$ are known in the literature~\cite{Vogt:2004mw,Wang:2019tgg}, which are listed in the supplemental material~\cite{sup.mat.} for reference. The objective of our calculations is to compute $\t f^{(2)}_{q/q''}$ and extract $C^{(2)}_{q q''}$. In the perturbative calculations at this order, $\t f^{(2)}_{q/q''}$ contains only IR divergences, 
which can be expressed as $1/\epsilon_{\rm IR}$ in the dimensional regulation. According to the factorization theorem, the IR divergences in $\t f_{q/q''}^{(2)}$ will be cancelled by that from the right hand side of Eq.~(\ref{match.scheme.simple}). In particular, the $1/\epsilon_{\rm IR}^2$ term will be cancelled by the last term and the $1/\epsilon_{\rm IR}$ by the second and last term. After these cancellations, we are left with a finite term, which will be the matching coefficient at this order.

To obtain the two loop contributions in Fig.~\ref{tab:FDsea}, some calculation techniques are employed and we will take the subdiagram (c) as an example.  In the  covariant $R_\xi$ gauge, Fig.~\ref{tab:FDsea}(c) contributes
\begin{widetext}
\begin{align}
\t f^{(2)}_{q/q''}(y, \frac{ p^z}{\mu})|_{Fig.~\ref{tab:FDsea}(c)}=&\mu^{4\epsilon}\int\int\frac{d^{4-2\epsilon} k_1}{(2\pi)^{4-2\epsilon}}\frac{d^{4-2\epsilon} k_2}{(2\pi)^{4-2\epsilon}}\bar u(p)(-ig T^a\gamma_{\mu_2})\frac{i}{\slashed p+\slashed k_2} (-igT^b\gamma_{\mu_4})u(p)\frac{-i}{k_2^2}\left(g^{\mu_1\mu_2}-(1-\xi)\frac{k_2^{\mu_1}k_2^{\mu_2}}{k_2^2}\right)\nonumber\\
&\times (-1)\mathrm{Tr}\left[\gamma^t \frac{i}{\slashed k_1}(-ig T^b\gamma_{\mu_3})\frac{i}{\slashed k_1+\slashed k_2} (-igT^a\gamma_{\mu_1})\frac{i}{\slashed k_1} \right]\frac{-i}{k_2^2}\left(g^{\mu_3\mu_4}-(1-\xi)\frac{k_2^{\mu_3}k_2^{\mu_4}}{k_2^2}\right)\frac{\delta(y-\frac{k_1^z}{p^z})}{
4N_c p^t}.
\label{ampc}
\end{align}
\end{widetext}
In the axial gauge $A^z=0$, one can also easily write down the contribution by replacing the gluon propagators
into $(-i)/k_2^2[g^{\mu_i\mu_j}-(n^{\mu_i}k_2^{\mu_j}+n^{\mu_j}k_2^{\mu_i})/(n\cdot k_2)+n^2k_2^{\mu_i}k_2^{\mu_j}/(n\cdot k_2)^2]$. Only subdiagram-(c) and subdiagram-(e) in Fig.~\ref{tab:FDsea} contribute in the  axial gauge, where the contribution of Fig.~\ref{tab:FDsea}(e)
can be obtained from Eq.~(\ref{ampc}) by the replacement of $p\to-p$ and the replacement of gluon propagators in axial gauge.

 To use the integration-by-parts (IBP) technique~\cite{Smirnov:2014hma} and reduce all the involved  tensor integrals into a set of integrals called master integrals, we use the identity
 \begin{align}
\delta(y-\frac{k_1^z}{p^z})=&\frac{p^z}{2\pi i}\left(\frac{1}{k_1^z-y p^z-i0}-\frac{1}{k_1^z-y p^z+i0}\right).
\end{align}
Method of differential equations \cite{Kotikov:1990kg, Kotikov:1991pm,Henn:2013pwa} are applied to calculate those master integrals. All the analytic expression of  master integrals  are given in Ref.~\cite{Chen:2020iqi} by the current authors. As a specific example, all the Feynman integrals from
Eq.~(\ref{ampc}) can be expressed by the first family of integrals listed in  Ref.~\cite{Chen:2020iqi}.
We have checked that the final results in  covariant and axial gauges are consistent with each other.

As mentioned above, there is no UV divergence  in  flavor non-diagonal quark quasidistributions and thus it is not necessary to perform the renormalization in modified minimal subtraction scheme. All soft divergences are  also cancelled.
The collinear divergences in $0<y<1$ region contain $1/\epsilon_{\rm IR}^2$ and $1/\epsilon_{\rm IR}$:
\begin{align}
\t f^{(2)}_{q/q''}(y, \frac{ p^z}{\mu})|_{div.,0<y<1}=&\frac{1}{ \epsilon_{\mathrm{IR}}^2}\Gamma_2(y)+\frac{1}{\epsilon_{\rm IR}}\Gamma_1(y)\nonumber\\&+\frac{2}{\epsilon_{\rm IR}}\Gamma_2(y)\log (\frac{\mu^2}{ {p^z}^2}) \ ,
\label{div01}
\end{align}
where $\Gamma_1$ and $\Gamma_2$ are defined as
\begin{widetext}
\begin{eqnarray}
\Gamma_1(y)&=&{2T_F C_F}[\frac{\left(4 y^3+3 y^2-3 y-6 (y+1) y \log (y)-4\right)}{3 y}\log (2)+(y-1) \text{Li}_2(-y)+(y+1) \text{Li}_2(y)\nonumber\\&~~&+\frac{(y+1) \left(8 y^2+y-16\right) \log (y)}{12 y}+\frac{y \left(3 \pi ^2
   (y-1)-y (8 y+57)+9\right)-10}{18 y}-\frac{1}{4} (3 y+7) \log ^2(y)\nonumber\\&~~&+\frac{(y-1) (y (4 y+7)+4) \log
   (1-y)}{6 y}-\frac{(y+1) (y (4 y-7)+4) \log (y+1)}{6 y}],\label{gamma1}
\end{eqnarray}
\end{widetext}
\begin{align}
\Gamma_2(y)=-\frac{T_{F}C_{F}\left(4 y^{3}+3 y^{2}-3 y-6(y+1) y \log (y)-4\right)}{6 y}. \label{gamma2}
\end{align}

These divergences will be cancelled by two parts in Eq.~(\ref{match.scheme.simple}): one from the divergences in the convolution of $C^{(1)}_{qg}\otimes f^{(1)}_{gq''}$ and the other from the divergences in $f^{(2)}_{qq''}$.  Both of them are listed  in the supplemental material~\cite{sup.mat.}  with know results of $C^{(1)}_{qg}$ and $f^{(1)}_{g/q''}$ at one-loop order, and $f^{(2)}_{qq''}$ at two loop. 
For the collinear divergences in $-1<y<0$ region, one can obtain from Eq.~(\ref{div01}) and do the replacement $y\to -y$ and add a prefactor -1.
Note that one should do the $\log (p(y))\to\log(p(y)^2)/2$ replacement at first to ovoid to produce the imaginary part. The IR cancellation is similar.

\begin{equation}
\begin{aligned}
\t f^{(2)}_{q/q''}(y, \frac{ p^z}{\mu},\epsilon_{\mathrm{IR}})|_{div.,y>1}&=\frac{1}{ \epsilon_{\mathrm{IR}}}\Gamma_1'(y)\ ,\\
\t f^{(2)}_{q/q''}(y, \frac{ p^z}{\mu},\epsilon_{\mathrm{IR}})|_{div.,y<-1}&=-\frac{1}{ \epsilon_{\mathrm{IR}}}\Gamma_1'(-y)\ ,
 \end{aligned}
\end{equation}
where
\begin{align}
\Gamma_1'(y)=&-\frac{T_F C_F}{3 y }[6 (y-1) y \text{Li}_2\left(-\frac{1}{y}\right)+22 y
\nonumber\\&-6 (y+1) y \text{Li}_2\left(\frac{1}{y}\right)+\left(3-4
   y^2\right) y\log \left(\frac{y^2}{y^2-1}\right)\nonumber\\
   &+\left(4-3 y^2\right) \log \left(\frac{y+1}{y-1}\right)] \ .\label{gamma1p}
\end{align}
These divergences are cancelled by the convolution of $C^{(1)}_{qg}\otimes f^{(1)}_{g/q''}$ as indicated in Eq.~(\ref{match.scheme.simple}). Again, we list the result in the supplemental material~\cite{sup.mat.}.   One can also see the two regions of $y>1$ and $y<-1$ are related by the symmetry of $y\to -y$ and an opposite sign.

The nontrivial cancellation of the IR divergences discussed above is an important demonstration of the LaMET factorization. This also provides a cross check of our final result on the matching coefficient, which will be presented in the next section.

We would like to emphasize a number of points before we close this sections. First, the complete cancellation of the collinear divergence depends on the factorization formula for this channel, see, Eq.~(\ref{match.scheme.simple}), including the different terms contributing from the right hand side. Second, it also depends on the exact results of lower order perturbative contributions. For example, the scale dependent term in the one-loop matching $C_{gq}^{(1)}$ (see in the supplemental material~\cite{sup.mat.} ) plays a crucial role to demonstrate the complete cancellation in the above equations. This emphasizes the importance of a consistent subtraction scheme in the perturbative calculations of Quasi-PDFs and the matching coefficients. Finally, our example of the non-diagonal quark-quark channel shall provide important guideline for future developments on computing the Quasi-PDFs at two-loop order.

 \textit{Matching Coefficient at Two-loop Order.}---
The matching coefficient $C_{qq''}^{(2)}$ is obtained by expanding both sides of Eq.~(\ref{match.scheme.simple}) to ${\cal O}(\epsilon^0)$ order. Because of all the divergences between them have been cancelled explicitly as shown in the previous section, it is straightforward to carry out the calculations for the finite parts.

First, let us show the result of NNLO matching coefficient $C^{(2)}_{qq''}$ in the region of $x>1$
\begin{widetext}
\begin{align}
C^{(2)}_{qq''}\Big(y, \frac{ p^z}{\mu} \Big)|_{y>1}&= T_{F}C_{F} \left[\Gamma_1'(y)\log \left(\frac{\mu^2}{{p^z}^2}\right)+\left(2-\frac{8 y^2}{3}\right) \log ^2(y)-\frac{4}{9} \left(10 y^2+9\right) \log
   (y)\right.\nonumber\\&~~\left.-\frac{4}{3} \log (2) \left(\left(4 y^2-3\right) \log (y)-11\right)+\frac{4
   \log ^3(y)}{3}-\frac{106}{9}+g_1(y)\right],
 \end{align}

where $\Gamma_1'(y)$ has been defined in Eq.~(\ref{gamma1p}) and  $g_1(y)$ is defined as
\begin{align}
g_1(y)&=4 (y+1) \text{Li}_3\left(\frac{1}{1-y}\right)+2 (y+1)
   \text{Li}_3\left(\frac{1}{y}\right)-\frac{(y+1)
   \text{Li}_2\left(\frac{1}{y}\right) \left(y (8 y-5)+6 y \log \left(4
   (y-1)^2\right)+8\right)}{3 y}\nonumber\\&~~+\frac{\left(8 (y-1) (y
   (5 y-16)+5)-9 y (y+1) \log ^2\left(y^2\right)\right) \log
   \left((y-1)^2\right)}{36 y}+\left(\frac{4 y^2}{3}+y-\frac{4}{3 y}-1\right) \log
   (2) \log \left((y-1)^2\right)\nonumber\\&~~+\frac{\left(3 y (y+1) \log \left(y^2\right)+(y-1)
   (y (4 y+7)+4)\right) \log ^2\left((y-1)^2\right)}{12 y}-\frac{1}{12} (y+1) \log ^3\left((y-1)^2\right)+[y\to -y]\ .
 \end{align}

Similarly,
we can obtain matching coefficient in the $0<y<1$ region

\begin{align}
&C^{(2)}_{qq''}\Big(y, \frac{ p^z}{\mu} \Big)|_{0<y<1}\nonumber\\&=\Gamma_2(y)\log^2\left(\frac{\mu^2}{{p^z}^2}\right)+\left(\Gamma_1(y)-P_{q/q''}^{(1)}(y)\right)
\log\left(\frac{\mu^2}{{p^z}^2}\right)+T_{F} C_{F} \left[\frac{2 (y-2) (y+1)^2 \text{Li}_2(-y)}{3 y}+\frac{1}{6} (y+13) \log
   ^3(y)\right.\nonumber\\&~~+\left(y^2+\frac{7
   y}{4}+\frac{8}{3 y}+\frac{15}{4}\right) \log ^2(y)+\log ^2(2) \left(-\frac{8
   y^2}{3}-2 y+\frac{8}{3 y}+4 (y+1) \log (y)+2\right)+\frac{250 y^2}{27}\nonumber\\&~~+\frac{\pi ^2 \left(y
   \left(4 y^2-3 y+6 (1-y) \log (2)-3\right)+6 y \log (y)-2\right)}{9 y}+4 (y+1)
   \zeta (3)-\frac{419 y}{18}+\frac{56}{27 y}\nonumber\\&~~+\frac{\log (2) (-4 y (2 y (5 y-21)+9)+6 \log (y) (3 y (y+3)+3 y
   (y+3) \log (y)+8)+40)}{9 y}+\frac{1}{6}\nonumber\\&~~\left.+\frac{\left(-y (4 y (14 y+9)+15)+6 (y-2) (y+1)^2 \log (y+1)+40\right)
   \log (y)}{9 y}+g_2(y)\right],
 \end{align}
where $P_{q/q''}^{(1)}(y)$ being the two loop splitting function
\begin{equation}
P_{q/q''}^{(1)}(y)=\frac{T_F C_F}{2}\left[\frac{20}{9y}-2+6y-\frac{56y^2}{9}+\left(1+5y+\frac{8y^2}{3}\right)\log (y)-(1+y)\log^2 (y)\right]\theta(y)\theta(1-y).
\end{equation}
and $g_2(y)$ is defined as
\begin{align}
g_2(y)&=\text{Li}_2(y) \left(-\frac{10 y^2}{3}-y-\frac{4}{y}-4 (y+1) \log (2-2
   y)+1\right)-4 (y+1) \text{Li}_3(1-y)-2 (y+1) \text{Li}_3(y)\nonumber\\&~~-\frac{(y-1)
   \left(3 \left(y^2+y-2\right) \log \left(y^2\right)+4 (y (5 y-16)+5)\right) \log
   (1-y)}{9 y}+\frac{2}{3} \pi ^2 (y+1) \log (1-y)\nonumber\\&~~+\left(1-\frac{4
   y^2}{3}-y+\frac{4}{3 y}-2 (y+1) \log (y)\right) \log ^2(1-y)-\frac{2 (y-1) (y (4 y+7)+4)
   \log (2) \log (1-y)}{3 y}-[y\to -y].
 \end{align}
\end{widetext}

\begin{figure}[th]
\begin{center}
\includegraphics[width=0.45\textwidth]{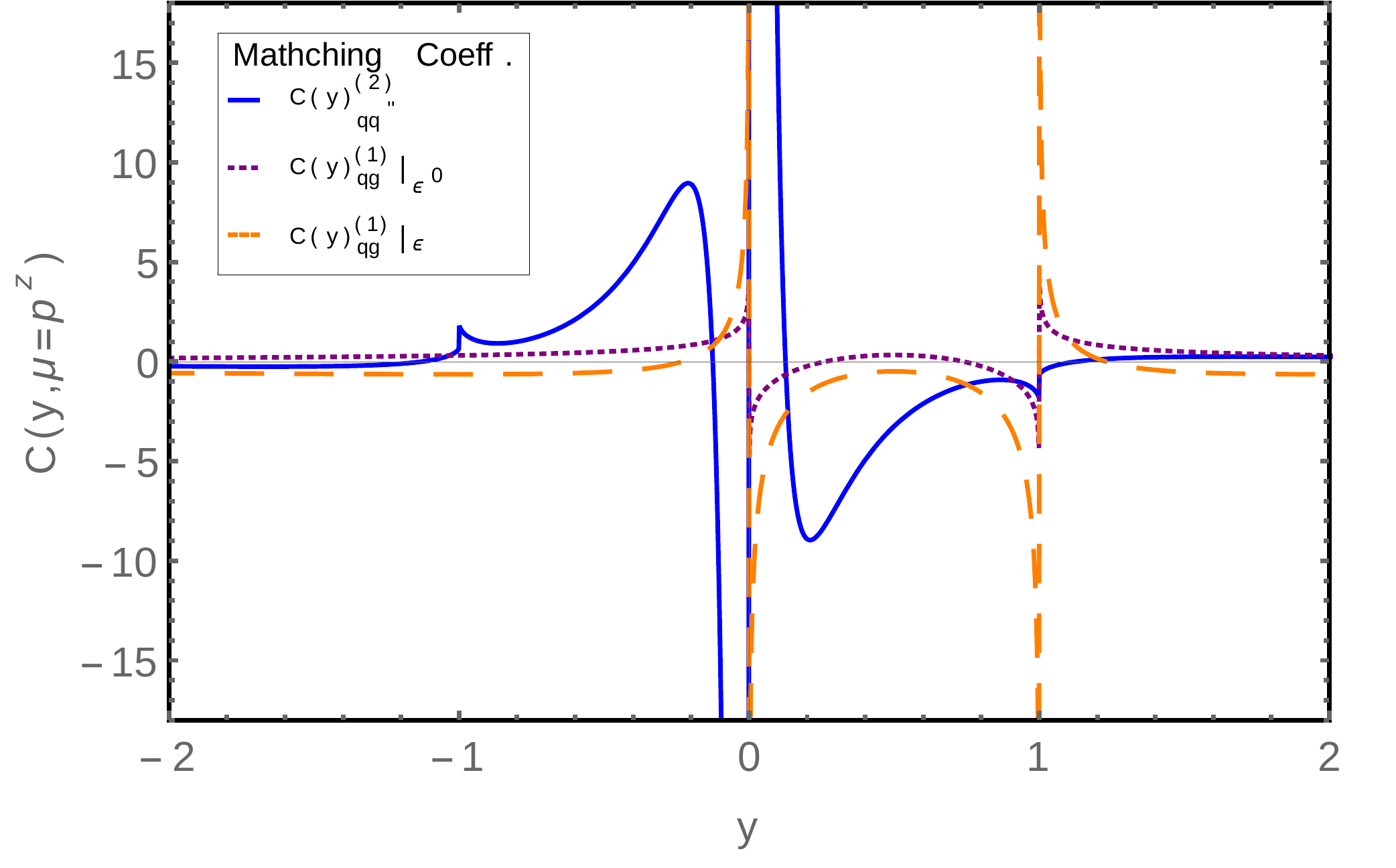}
\caption{ Distributions of matching coefficients of $C^{(2)}_{qq''}$ and $C^{(1)}_{qg}$ as a function of momentum fraction $y$, where we adopt the scale $\mu=p^z$. }\label{fig:c2plot}
\end{center}
\end{figure}

In the end, the NNLO matching coefficient $C^{(2)}_{qq''}$ in $y<0$ can be obtained by replacing  $y\to -y$  and adding an overall minus sign.
It is interesting to investigate the asymptotic behaviour of the matching coefficients at infinity points. Up to $\mathcal{O}(\epsilon)$, we have, for example,
\begin{align}
C^{(1)}_{qg}\left(y,\frac{p^z}{\mu}\right)|_{y\to +\infty}=&\frac{2}{3y}T_F \left[1+2 \log (2 y)\epsilon\right],\nonumber\\
C^{(1)}_{qg}\left(y,\frac{p^z}{\mu}\right)|_{y\to -\infty}=&-\frac{2}{3y}T_F \left[1+2 \log (-2 y)\epsilon\right] \ .
\end{align}
These will lead to a logarithmic divergence when performing the integration of $C^{(1)}_{q g}\Big(\frac{y}{x}, \frac{|x| p^z}{\mu} \Big)\otimes f^{(1)}_{g/q''}(x, \epsilon_{\mathrm{IR}})$ at infinity points. However, they do cancel between the integration at positive and negative infinity points and thus we do not need to add prescriptions  to the divergence  at  the integration of infinity points. For the NNLO matching coefficients, we have
\begin{align}
&C^{(2)}_{qq''}\left(y,\frac{p^z}{\mu}\right)|_{y\to +\infty}
\nonumber\\&~~=\frac{4}{27y^2} T_F C_F
\left[-6\left(\frac{\mu^2}{{p^z}^2}\right)+12 \log (2 y)-7\right],\end{align}
\begin{align}
&C^{(2)}_{qq''}\left(y,\frac{p^z}{\mu}\right)|_{y\to -\infty}\nonumber\\&~~=-\frac{4}{27y^2} T_F C_F\left[-6\left(\frac{\mu^2}{{p^z}^2}\right)+12 \log (-2 y)-7\right].
\end{align}
From these, one can see $C^{(2)}_{qq''}\left(y,\frac{p^z}{\mu}\right)$ have better asymptotic behaviours at infinity points than  $C^{(1)}_{qg}\left(y,\frac{p^z}{\mu}\right)$ and there will be no divergences at $\pm \infty$ if one carries out, e.g., the integral of $C^{(2)}_{q q''}\Big(\frac{y}{x}, \frac{|x| p^z}{\mu} \Big)\otimes f^{(1)}_{q''/q''}(x, \mu)$ at NNNLO matching.

\begin{figure}[th]
\begin{center}
\includegraphics[width=0.45\textwidth]{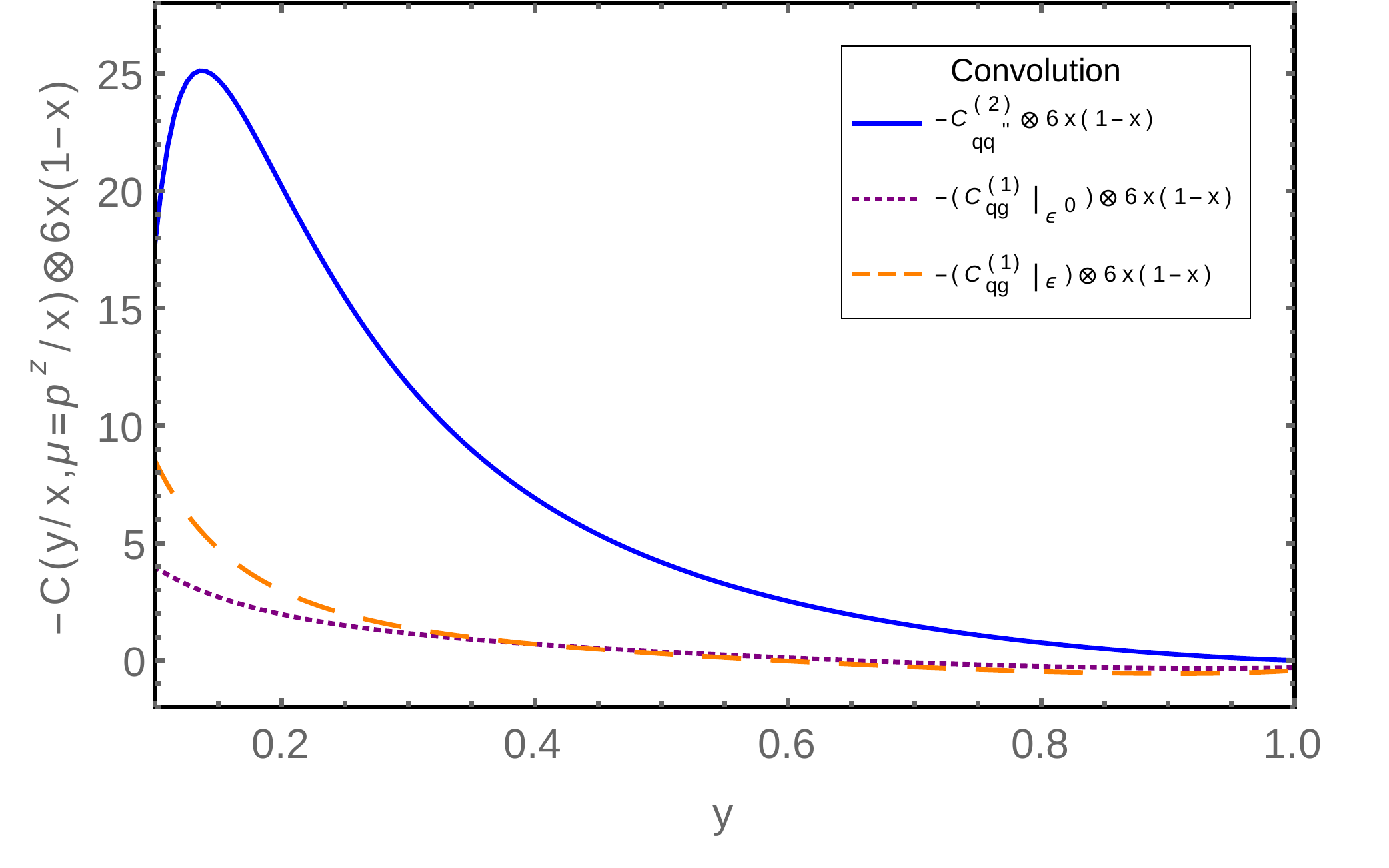}
\includegraphics[width=0.45\textwidth]{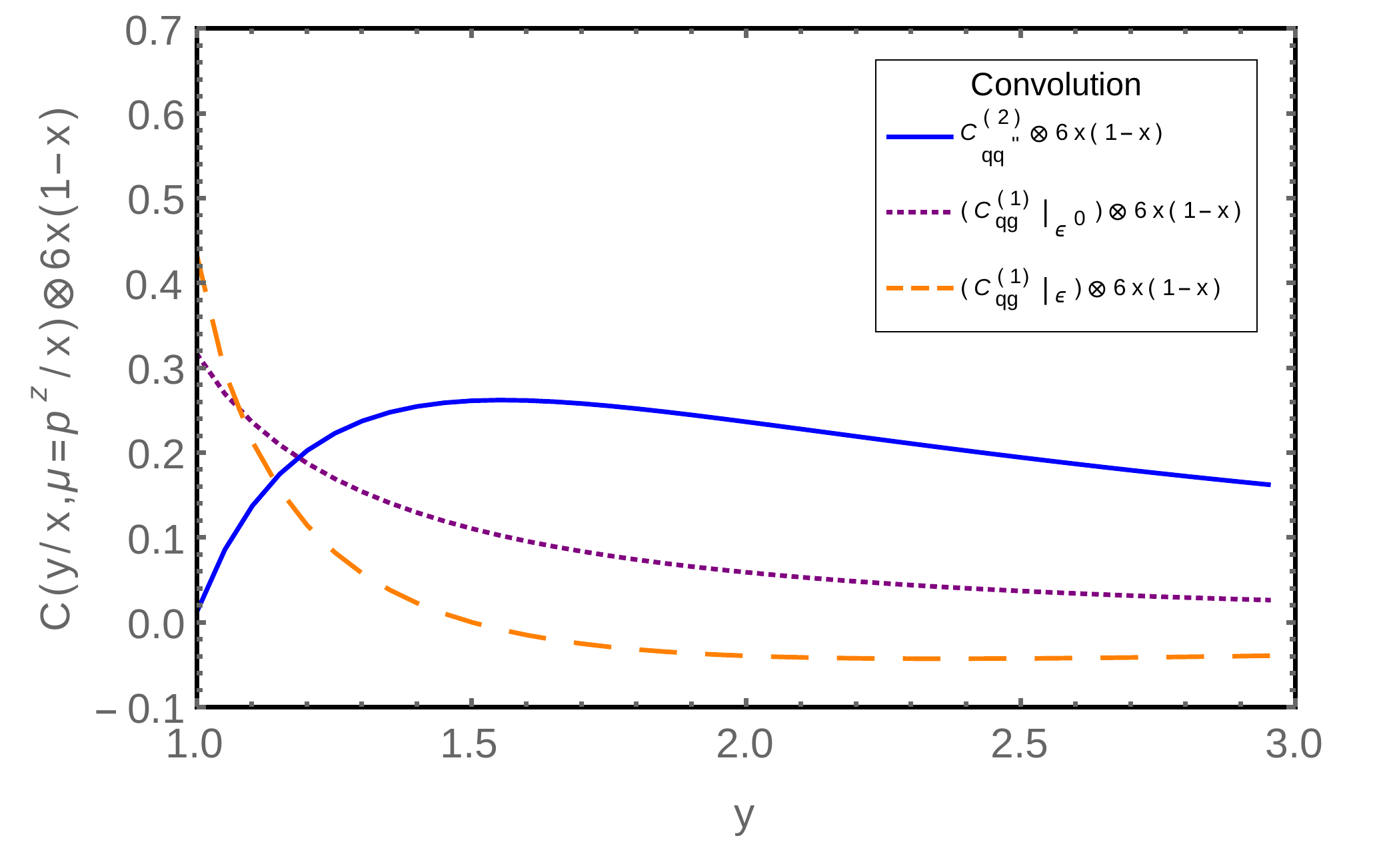}
\caption{ Distributions of the convolution of  $C^{(n)}_{ij}\left(\frac{y}{x},\frac{p^z}{\mu}\right)\otimes 6x(1-x)$ as a function of momentum fraction $y$, where we assume there is a toy model of $f(x)=6x(1-x)$ and adopt the scale $\mu=p^z/x$. Therein the up one is for $0.1<y<1$ and the bottom one is for $1<y<3$. }\label{fig:c2plotcon}
\end{center}
\end{figure}

We plot the distributions of matching coefficients of $C^{(2)}_{qq''}$ and $C^{(1)}_{qg}$ as a function of momentum fraction $y$ in Fig.~\ref{fig:c2plot}. Therein the renormalization scale is adopted as $\mu=p^z$ and the lattice realization of $p^z$ is in several GeV currently. From it, $C^{(2)}_{qq''}$ has a different shape compared with others. Assuming the parameterization form of light cone PDFs as the simplest one $a x^b (1-x)^c$, we can  test the convolution between matching coefficients and light cone PDFs. So we also plot the convolution of  $C^{(n)}_{ij}\left(\frac{y}{x},\frac{|x|p^z}{\mu}\right)\otimes 6x(1-x)$  in Fig.~\ref{fig:c2plotcon} as a toy model.   Note that $C^{(1)}_{qg}$ is separate into $C^{(1)}_{qg}\left(x,\frac{p^z}{\mu}\right)|_{\epsilon^0}$ and $C^{(1)}_{qg}\left(x,\frac{p^z}{\mu}\right)|_{\epsilon}$, where $C^{(1)}_{qg}\left(y,\frac{p^z}{\mu}\right)|_{\epsilon^0}$ does not depend on the renormalization scale in non-physical region, but $C^{(1)}_{qg}\left(y,\frac{p^z}{\mu}\right)|_{\epsilon}$  depends on the renormalization scale in all the region.
$C^{(2)}_{qq''}$ has double logarithms as $\Gamma_2 \log^2 (\frac{\mu^2}{{p^z}^2})$ in physical region, while single logarithms as $\Gamma_1' \log (\frac{\mu^2}{{p^z}^2})$ in non-physical region.

 \textit{Conclusion.}---
In summary, we have presented a next-to-next-to-leading order calculation of the quasi parton distribution functions for the  flavor non-diagonal quark contributions $\t f^{(2)}_{q/q''}(y, \frac{ p^z}{\mu})$ in $-\infty<y<\infty$. We have demonstrated the LaMET factorization at this order. The matching coefficient is derived under the modified minimal subtraction scheme.  These results shall be directly employed to investigate the sea quark contributions in both nonsinglet and singlet quark distributions at NNLO. This will stimulate further developments toward a complete calculation of Quasi-PDFs at two-loop order and the associated matching coefficients.

 \textit{Acknowledgements}---We thank F. Yuan for the helps to solve the convolution integrals, the valuable advices on our manuscript and all  valuable discussions during the work. We thank X. Ji, Y.-S. Liu,  J. Wang, L.-L. Yang, Y. Zhao for valuable discussions.
LBC is supported by the National Natural Science Foundation of China (NSFC) under the grant No.~11805042. WW is supported by NSFC under grants No.~11735010, 11911530088,  by Natural Science Foundation of Shanghai under grant No. 15DZ2272100. RLZ is supported by NSFC under grant No.~11705092, by Natural Science Foundation of Jiangsu under Grant No.~BK20171471, by China Scholarship Council under Grant No.~201906865014 and partially supported by the U.S. Department of Energy, Office of Science, Office of Nuclear Physics, under contract number DE-AC02-05CH11231.

\begin{widetext}
\appendix

\section{Matching procedure and IR divergences in Quasi PDFs}

Generically the matching between quasi and light-cone PDFs is given as
\begin{equation}
\begin{aligned}
&\t f^{(0)}_{i/k}(y, \frac{ p^z}{\mu},\epsilon_{\mathrm{IR}})=C^{(0)}_{ij}\left(\frac{ y}{x}, \frac{ |x| p^z}{\mu}\right)\otimes
  f^{(0)}_{j/k}(x,\epsilon_{\mathrm{IR}}),\\
&\t f^{(1)}_{i/k}(y, \frac{ p^z}{\mu},\epsilon_{\mathrm{IR}})=C^{(1)}_{ij}\left(\frac{ y}{x}, \frac{ |x| p^z}{\mu}\right)\otimes
  f^{(0)}_{j/k}(x,\epsilon_{\mathrm{IR}})+C^{(0)}_{ij}\left(\frac{ y}{x}, \frac{ |x| p^z}{\mu}\right)\otimes
  f^{(1)}_{j/k}(x,\epsilon_{\mathrm{IR}}),\\
&\t f^{(2)}_{i/k}(y, \frac{ p^z}{\mu},\epsilon_{\mathrm{IR}})=C^{(2)}_{ij}\left(\frac{ y}{x}, \frac{ |x| p^z}{\mu}\right)\otimes
  f^{(0)}_{j/k}(x,\epsilon_{\mathrm{IR}})+C^{(1)}_{ij}\left(\frac{ y}{x}, \frac{ |x| p^z}{\mu}\right)\otimes
  f^{(1)}_{j/k}(x,\epsilon_{\mathrm{IR}})+C^{(0)}_{ij}\left(\frac{ y}{x}, \frac{ |x| p^z}{\mu}\right)\otimes
  f^{(2)}_{j/k}(x,\epsilon_{\mathrm{IR}}).
\end{aligned}
\end{equation}
 For the non-diagonal quark-quark splitting at two-loop order, we need to consider the contributions from $C_{qg}^{(1)}$ and $f_{g/q}^{(1)}$ for the second term, $C_{qq}^{(0)}$ and $f_{q''/q}^{(2)}$ for the third term, see Section ``Factorization at Two-loop Order'' in the main text. $C_{qq}^{(0)}$ is a Delta function. In the following, we list other terms for the reference. To extract the matching coefficient $C_{qq''}$, we will keep some of the terms up to ${\cal O}(\epsilon)$.

First, the matching coefficient $C_{q/g}^{(1)}$ can be written as, 
\begin{align}
C^{(1)}_{qg}\left(y,\frac{p^z}{\mu}\right)&=  C^{(1)}_{qg}\left(y,\frac{p^z}{\mu}\right)|_{\epsilon^0}+\epsilon   C^{(1)}_{qg}\left(y,\frac{p^z}{\mu}\right)|_{\epsilon},
\end{align}
up to ${\cal O}(\epsilon)$ order. The leading term has been calculated in Ref.~\cite{Wang:2019tgg}
\begin{eqnarray}
C^{(1)}_{qg}\left(y,\frac{p^z}{\mu}\right)|_{\epsilon^0}&=T_F\left\{ \begin{array}{ll}\left(-2 y^2+2 y-1\right) \log \left(\frac{y-1}{y}\right)-2 y+1,\;\;&y>1\\
-\left(2 y^2-2 y+1\right) \log \left(\frac{\mu ^2}{4 y(1-y)  {p^z}^2}\right)-6 y^2+6 y-1,\;\;&0<y<1\\
\left(2 y^2-2 y+1\right) \log \left(\frac{y-1}{y}\right)+2 y-1,\;\;&y<0.\end{array}\right.\label{eq:cqgdef}\end{eqnarray}
We have calculated the ${\cal O}(\epsilon)$ as well,
\begin{eqnarray}\begin{aligned}
&C^{(1)}_{qg}\left(y,\frac{p^z}{\mu}\right)|_{\epsilon}/T_F\\&=\left\{ \begin{array}{ll}\left(-\left(2 y^2-2 y+1\right) \log \left(\frac{y-1}{y}\right)-2 y+1\right) \left(1+\log \left(\frac{\mu
   ^2}{4 {p^z}^2}\right)\right)+\left(2 y^2-2 y+1\right) \left(\log ^2(y-1)-\log ^2(y)\right)\\+\left(-4 y^2+6 y-1\right) \log
   \left(\frac{y-1}{y}\right)+2 (2 y-1) (\log (y)-1),\;\;&y>1\\
-\left(2 y^2-2 y+1\right) \left(\frac{1}{2} \log ^2\left(\frac{\mu ^2}{4 {p^z}^2}\right)+\log \left(\frac{\mu ^2}{4
   {p^z}^2}\right)+\frac{\pi ^2}{4}+2\right)-\left(2 y^2-2 y+1\right) \left(\log ^2(1-y)+\log ^2(y)\right)\\+\left(\left(2 y^2-2 y+1\right) \log (-(y-1) y)-4 (y-1) y\right) \left(\log
   \left(\frac{\mu ^2}{4 {p^z}^2}\right)+1\right)+\left(4
   y^2-6 y+1\right) \log ((1-y) y)\\-6 (y-1) y+2 (2 y-1) \log (y)-1,\;\;&0<y<1,\\
\left(\left(2 y^2-2 y+1\right) \log \left(\frac{y-1}{y}\right)+2 y-1\right) \left(1+\log \left(\frac{\mu
   ^2}{4 {p^z}^2}\right)\right)-\left(2 y^2-2 y+1\right) \left(\log ^2(1-y)-\log ^2(-y)\right)\\-\left(-4 y^2+6 y-1\right) \log
   \left(\frac{y-1}{y}\right)-2 (2 y-1) (\log (-y)-1),\;\;&y<0.\end{array}\right.
\end{aligned}\end{eqnarray}

Now, we turn to the light-cone parton distribution functions $f_{g/q}^{(1)}$ and $f_{q/q''}^{(2)}$ in  the $\overline{\rm MS}$ scheme. In general, we have the light-cone distribution functions as~\cite{Luo:2019hmp}
\begin{eqnarray}
f_{i / j}^{(0)}(x)&=&\delta_{ij}\delta(1-x)\ ,\\
f_{i / j}^{(1)}(x)&=&-\frac{P_{i j}^{(0)}(x)}{\epsilon_{\mathrm{IR}}}\ , \\
f_{i / j}^{(2)}(x)&=&\frac{1}{ 2\epsilon_{\mathrm{IR}}^{2}}\left[\sum_{k} P_{i k}^{(0)}(x) \otimes P_{k j}^{(0)}(x)+\beta_{0} P_{i j}^{(0)}(x)\right]-\frac{P_{i j}^{(1)}(x)}{\epsilon_{\mathrm{IR}}} \ .
\end{eqnarray}
In our case, we need the following leading order Altarelli-Parisi splitting functions~\cite{Altarelli:1977zs}
\begin{equation}
\begin{aligned}
&P_{g q}^{(0)}(x)= C_{F} \frac{1+(1-x)^{2}}{y}, \\
&P_{q g}^{(0)}(x)= T_{F} [x^{2}+(1-x)^{2}].
\end{aligned}
\end{equation}
From the above equations, we readily have the expression for $f_{g/q}^{(1)}=-\frac{1}{\epsilon_{\rm IR}}P^{(0)}_{gq}$. Combining this with $C_{qg}^{(1)}$, we will be able to obtain the divergence contributions from the second term of Eq.~(6)  in the main text. In the region of $y>1$,
one has
\begin{equation}
\begin{aligned}
&C^{(1)}_{qg}\left(\frac{ y}{x}, \frac{ |x |p^z}{\mu}\right)|_{\epsilon^0}\otimes
  f_{g / q}^{(1)}(x)\\
  &=-\frac{1}{ \epsilon_{\mathrm{IR}}}\int^{+\infty}_{-\infty}dy_1 \int^{1}_{-1}dx
 C^{(1)}_{qg}\left(y_1, \frac{ |x |p^z}{\mu}\right)P_{gq}^{(0)}(x)\delta(y-y_1 x)\\
 &=-\frac{1}{ \epsilon_{\mathrm{IR}}}\left[\int^{+\infty}_{y}dy_1 \frac{1}{y_1}
 C^{(1)}_{qg}\left(y_1, \frac{ y p^z}{y_1\mu}\right)P_{gq}^{(0)}(\frac{y}{y_1})+\int^{-y}_{+\infty}dy_1 \frac{-1}{y_1}
 C^{(1)}_{qg}\left(y_1, -\frac{ y p^z}{y_1\mu}\right)(-1)P_{gq}^{(0)}(-\frac{y}{y_1})\right]\\
  &=-\frac{1}{ \epsilon_{\mathrm{IR}}}\int^{+\infty}_{y}dy_1 \frac{1}{y_1}
 \left[C^{(1)}_{qg}\left(y_1, \frac{ y p^z}{y_1\mu}\right)-C^{(1)}_{qg}\left(-y_1, \frac{ y p^z}{y_1\mu}\right)\right]P_{gq}^{(0)}(\frac{y}{y_1})\\
 &=-\frac{T_F C_F}{3 y \epsilon_{\mathrm{IR}}}\left[6 (y-1) y \text{Li}_2\left(-\frac{1}{y}\right)-6 (y+1) y \text{Li}_2\left(\frac{1}{y}\right)+\left(3-4
   y^2\right) y \log \left(\frac{y^2}{y^2-1}\right)\right.\\&\left.~~+\left(4-3 y^2\right) \log \left(\frac{y+1}{y-1}\right)+22 y\right] \ .
 \end{aligned}
\end{equation}
For $0<y<1$, on the other hand, we obtain
\begin{equation}
\begin{aligned}
&C^{(1)}_{qg}\left(\frac{ y}{x}, \frac{ |x | p^z}{\mu}\right)|_{\epsilon^0}\otimes
  f_{g / q}^{(1)}(x)\\
  &=-\frac{1}{ \epsilon_{\mathrm{IR}}}\int^{+\infty}_{-\infty}dy_1 \int^{1}_{-1}dx
 C^{(1)}_{qg}\left(y_1, \frac{ |x| p^z}{\mu}\right)P_{gq}^{(0)}(x)\delta(y-y_1 x)\\
 &=-\frac{1}{ \epsilon_{\mathrm{IR}}}\left[\int^{1}_{y}dy_1 \frac{1}{y_1}
 C^{(1)}_{qg}\left(y_1, \frac{ y p^z}{y_1\mu}\right)P_{gq}^{(0)}(\frac{y}{y_1})+\int^{-y}_{-1}dy_1 \frac{1}{y_1}
 C^{(1)}_{qg}\left(y_1, -\frac{ y p^z}{y_1\mu}\right)P_{gq}^{(0)}(-\frac{y}{y_1})\right.\\
 &\left.~~+\int^{+\infty}_{1}dy_1 \frac{1}{y_1}
 \left(C^{(1)}_{qg}\left(y_1, \frac{ y p^z}{y_1\mu}\right)-C^{(1)}_{qg}\left(-y_1, \frac{ y p^z}{y_1\mu}\right)\right)P_{gq}^{(0)}(\frac{y}{y_1})\right]\\
 &=\frac{T_F C_F}{ \epsilon_{\mathrm{IR}}}\left[-\frac{\log (\frac{\mu^2}{4 {p^z}^2}) \left(4 y^3+3 y^2-3 y-6 (y+1) y \log (y)-4\right)}{3 y}+2 (y-1)
   \text{Li}_2(-y)\right.\\
   &~~+2 (y+1) \text{Li}_2(y)+\frac{2 \left(4 y^3+6 y^2-3 y-4\right) \log (y)}{3 y}+\frac{\left(4 y^3+3
   y^2-3 y-4\right) \log (1-y)}{3 y}\\
   &\left.~~+\frac{\left(-4 y^3+3 y^2+3 y-4\right) \log (y+1)}{3 y}+\frac{1}{3} \left(\pi ^2
   (y-1)-2 y (6 y+5)\right)-2 (y+2) \log ^2(y)\right].
 \end{aligned}
\end{equation}
The results of $y<0$ can be obtained with the replacement $y\to -y$ and an overall minus sign.

To get $f_{q/q''}^{(2)}$, one needs to apply the convolution,
\begin{equation}
f(x) \otimes g(x)=\int_{0}^{1} d y \int_{0}^{1} dz f\left(y\right) g\left(z\right)  \delta\left(x-y z\right)=\int_{x}^{1} \frac{\mathrm{d} z}{z} f\left(\frac{x}{z}\right) g(z)=\int_{x}^{1} \frac{\mathrm{d} y}{y} f(y) g\left(\frac{x}{y}\right)\ ,
\end{equation}
to obtain the $1/\epsilon_{\mathrm{IR}}^{2}$-term,
\begin{equation}
\begin{aligned}
\frac{1}{ 2\epsilon_{\mathrm{IR}}^{2}}P_{q g}^{(0)}(x) \otimes P_{g q}^{(0)}(x)&=-\frac{1}{\epsilon_{\mathrm{IR}}^{2}}\frac{ T_{F}C_{F} \left(4 x^{3}+3 x^{2}-3 x-6(x+1) x \log (x)-4\right)}{6 x}\theta(x)\theta(1-x) \ .
\end{aligned}
\end{equation}
There is also a $1/\epsilon_{\mathrm{IR}}$-term in $f_{q/q''}^{(2)}$~\cite{Furmanski:1980cm,Moch:2004pa},
\begin{equation}
-\frac{P_{q/q''}^{(1)}(x)}{\epsilon_{\mathrm{IR}}}=-\frac{T_F C_F}{2\epsilon_{\mathrm{IR}}}\left[\frac{20}{9x}-2+6x-\frac{56x^2}{9}+\left(1+5x+\frac{8x^2}{3}\right)\log (x)-(1+x)\log^2 (x)\right]\theta(x)\theta(1-x).
\end{equation}
The above two contribute to the divergences in the third term of Eq.~(6)  in the main text.

\section{Finite terms in Quasi PDFs}
First, let us show the result for the region of $y>1$.
The finite part of $f_{q/q''}^{(2)}$ is given by
\begin{align}
\t f^{(2)}_{q/q''}(y, \frac{ p^z}{\mu})|_{y>1}
&=2\Gamma_1'(y)\log\left(\frac{\mu^2}{{p^z}^2}\right)+T_{F}C_{F}\left[\frac{4}{3} \left(4 y^2-3\right) \log ^2(y)
+\log (2) \left(\left(8-\frac{32 y^2}{3}\right) \log
   (y)+\frac{88}{3}\right)\right.\nonumber\\&~~\left.+\frac{8 \log ^3(y)}{3}+\frac{4}{9} \left(8 y^2-9\right) \log
   (y)-\frac{298}{9}+h_1(y)\right] \ ,
 \end{align}

where $\Gamma_1'(y)$ has been defined in Eq.~(13)  in the main text and $h_1(y)$ is defined as

\begin{align}
h_1(y)&=\text{Li}_2\left(\frac{1}{y}\right) \left(-\frac{8 y^2}{3}+3 y-\frac{16}{3 y}-4
   (y+1) \log \left(4 (y-1)^2\right)-1\right)+8 (y+1)
   \text{Li}_3\left(\frac{1}{1-y}\right)+6 (y+1)
   \text{Li}_3\left(\frac{1}{y}\right)\nonumber\\&~~+\frac{1}{4} \left(\frac{8 y^2}{3}+2 (y+1)
   \log \left(y^2\right)+2 y-\frac{8}{3 y}-2\right) \log
   ^2\left((y-1)^2\right)+\frac{2
   \left(4 y^3+3 y^2-3 y-4\right) \log (2) \log \left((y-1)^2\right)}{3
   y}\nonumber\\&~~-\frac{\left(8 y^3+57 y^2+\frac{9}{2} (y+1) y \log
   ^2\left(y^2\right)-75 y+10\right) \log \left((y-1)^2\right)}{9 y}+\frac{1}{4} \left(4-\frac{16 y^2}{3}\right) \log
   ^2\left(y^2\right)\nonumber\\&~~-\frac{1}{6} (y+1) \log ^3\left((y-1)^2\right)+[y\to -y]\ .
 \end{align}

Similarly, the finite part of $f_{q/q''}^{(2)}$ in $0<y<1$ is given by
\begin{align}
&\t f^{(2)}_{q/q''}(y, \frac{ p^z}{\mu})|_{0<y<1}\nonumber\\&= 2\Gamma_2(y)
   \log^2\left(\frac{\mu^2}{{p^z}^2}\right)+ 2\Gamma_1(y)\log \left(\frac{\mu^2}{{p^z}^2}\right)+T_{F}C_{F}\left[\frac{2 (y-2) (y+1)^2 \text{Li}_2(-y)}{3 y}+\frac{\pi ^2 (y (y
   (4 y+3)-3)-12)}{36 y}\right.\nonumber\\&~~+\frac{2
   (y-2) (y+1)^2 \log (y+1) \log (y)}{3 y}+\frac{2}{9} \log (2)
   \left(2 \left(8 y^2+57 y-3 \pi ^2 (y-1)+\frac{10}{y}-9\right)+9 (3 y+7) \log
   ^2(y)\right)\nonumber\\&~~+\frac{1}{12}
   \left(-20 y^2-27 y+\frac{64}{y}+69\right) \log ^2(y)+\frac{4}{3} \log ^2(2)
   \left(-4 y^2-3 y+\frac{4}{y}+6 (y+1) \log (y)+3\right)\nonumber\\&~~+\frac{1}{18} \left(32 y^2+72 y-3 \pi ^2
   (y-11)+\frac{80}{y}-30\right) \log (y)-\frac{2 (y+1) \left(8 y^2+y-16\right)
   \log (2) \log (y)}{3 y}\nonumber\\&\left.~~+\frac{1}{6} (9 y+37) \log ^3(y)+12 (y+1) \zeta (3)-\frac{2 y^2}{27}-\frac{635 y}{18}+\frac{56}{27 y}+\frac{1}{6}+h_2(y)\right]
   ,
 \end{align}
where $\Gamma_1$ and $\Gamma_2$ are defined in Eqs.~(10,11) in the main text, respectively, and $h_2(y)$ is defined as
\begin{align}
h_2(y)&=\text{Li}_2(y) \left(-\frac{10 y^2}{3}+3 y-\frac{20}{3 y}-8 (y+1) \log (2-2
   y)+1\right)-8 (y+1) \text{Li}_3(1-y)-6 (y+1) \text{Li}_3(y)\nonumber\\&~~+\left(-\frac{8 y^2}{3}-2
   y+\frac{8}{3 y}-4 (y+1) \log (y)+2\right) \log ^2(1-y)+\frac{4}{3} \left(-4
   y^2-3 y+\frac{4}{y}+3\right) \log (2) \log (1-y)\nonumber\\&~~+\frac{\left(-3
   \left(y^3-3 y+2\right) \log (y^2)+2y \left(6 \pi ^2 (y+1)+y (8
   y+57)-75\right)+20\right) \log (1-y)}{9 y}-[y\to -y] \ .
 \end{align}
Note that the imaginary parts of terms by $\log (-y)\log ^2(1+y)/2$ and $\text{Li}_3(1+y)$ from the $y\to -y$ transform are cancelled.

\end{widetext}

\end{document}